\documentclass[letterpaper,12pt]{sapm}

\usepackage{geometry}                 
\usepackage{hyperref}

\usepackage{booktabs}
\usepackage{graphicx}

\usepackage{amsmath}
\usepackage{amssymb}
\usepackage{amsthm}
\usepackage{cite,xcolor}

\usepackage{subfig}

\newcommand{\C}{\mathbb{C}}
\newcommand{\I}{\mathrm{Im}\,}
\newcommand{\R}{\mathrm{Re}\,}
\newcommand{\hz}{H_0^{(1)}}
\newcommand{\db}{{\rm d}}

\jname{STUDIES IN APPLIED MATHEMATICS}
\jvol{AA}
\jyear{YYYY}
\doi{10.1111/((please add article doi))}

\begin{document}

\title[On Wiener-Hopf method for surface plasmons: Diffraction from semi-infinite sheet]
  {On the Wiener-Hopf method for surface plasmons:\\ Diffraction from semi-infinite metamaterial sheet}

  \author[D. Margetis, M. Maier, M. Luskin]{%
    Dionisios Margetis, Matthias Maier, and Mitchell Luskin
    \thanks{Address for correspondence: Prof. Dionisios Margetis, Department of
      Mathematics, and Institute for Physical Science and Technology, and Center for Scientific Computation and Mathematical Modeling, University of Maryland, College Park, MD 20742, USA. E-mail: \url{dio@math.umd.edu}}}

  \affil{
    Dionisios Margetis\\
    Department of Mathematics,\\
    Institute for Physical Science and Technology,\\
    Center for Scientific Computation and Mathematical Modeling,\\
    University of Maryland,\\
    College Park, Maryland 20742, USA.\\
    \url{dio@math.umd.edu}}
  \affil{
    Matthias Maier\\
    School of Mathematics\\
    University of Minnesota\\
    Minneapolis, Minnesota 55455, USA.\\
    \url{msmaier@umn.edu}}
  \affil{
    Mitchell Luskin\\
    School of Mathematics\\
    University of Minnesota\\
    Minneapolis, Minnesota 55455, USA.\\
    \url{luskin@umn.edu}}

  \maketitle

  \begin{abstract}
    By formally invoking the Wiener-Hopf method, we explicitly solve a
    one-dimensional, singular integral equation for the excitation of a slowly
    decaying electromagnetic wave, called surface plasmon-polariton (SPP),
    of small wavelength on a semi-infinite, flat conducting sheet
    irradiated by a plane wave in two spatial dimensions. This setting is germane to
    wave diffraction by edges of large sheets of single-layer
    graphene. Our analytical approach includes: (i) formulation of a
    functional equation in the Fourier domain;   (ii) evaluation of a {\em
    split function}, which is expressed by a contour integral and is a key
    ingredient of the Wiener-Hopf factorization; and (iii) extraction of
    the SPP as a simple-pole residue of a Fourier integral. Our analytical
    solution is in good agreement with a finite-element numerical
    computation.

  \end{abstract}

\section{Introduction}
\label{sec:Intro}

Surface plasmon-polaritons (SPPs) are evanescent electromagnetic waves that propagate on the surface of conductors and arise from the coupling of the incident radiation with the electron plasma~\cite{Bludov13,Pitarke07}. The study of surface waves in the radio-frequency regime dates back to Sommerfeld~\cite{Sommerfeld1896,Sommerfeld1899}; for later advances, see~\cite{King92}. Recently, it has been realized that SPPs with novel features can be generated on atomically thick, conducting metamaterials, e.g., graphene, in the infrared spectrum~\cite{Castroneto09,Geim13}. The dielectric permittivity of these materials can have a negative real part at terahertz frequencies. Accordingly, slowly decaying SPPs may emerge, confined near interfaces with a wavelength much smaller than the free-space wavelength. These waves form parts of solutions to boundary value problems for Maxwell's equations~\cite{Bludov13,Margetis16}.

In this article, we solve a prototypical scattering problem in order to analytically show how the SPP is excited via wave diffraction by the edge of a flat sheet with finite conductivity in the two-dimensional (2D) space, $\mathbb{R}^2$. By using boundary conditions of a resistive half plane\footnote{The terms ``conducting'' and ``resistive'' sheet or half plane are used interchangeably. The ``top Riemann sheet'', which is invoked below for a particular branch of a multiple-valued function in the complex plane, should not be confused with the physical sheet causing wave diffraction.} and the Sommerfeld radiation condition for Maxwell's equations with an $e^{-i\omega t}$ time dependence ($\omega>0$, $i^2=-1$), we state the problem in the form (see Appendix~\ref{app:BVP})\footnote{The placement of the one-dimensional Helmholtz operator, $(\db^2/\db x^2)+k^2$, outside the integral of~\eqref{eq:integro-diff} affords a kernel, $\mathcal K$, that is (logarithmically) {\em integrable} through $x=x'$;
cf.~Pocklington's integral equation for wire antennas~\cite{King-book,Rynne92}.
The problem can be stated in alternate forms, e.g., via direct application of the Fourier transform to a boundary value problem~\cite{Noble-book}, which circumvents~\eqref{eq:integro-diff}.}
\begin{align}\label{eq:integro-diff}
u(x)&=u^{\rm in}(x)+\frac{i\varsigma}{k}\Biggl(\frac{\db^2}{\db x^2}+k^2\Biggr)\int_0^\infty \mathcal K(x-x')\,u(x')\,\db x'~,\quad x>0~;\nonumber\\
& \mathcal K(x)=\frac{i}{4}\hz(k|x|)~,\quad -\infty<x<\infty~.
\end{align}
The kernel, $\mathcal K$, of integral equation~\eqref{eq:integro-diff} is expressed in terms of the first-kind, zeroth-order Hankel function, $\hz$~\cite{CourantHilbert}, which comes from the Green function for the scalar Helmholtz equation on $\mathbb{R}^2$~\cite{Colton83}.
Physically,~\eqref{eq:integro-diff} accounts for the continuity of the tangential electric field and a jump of the tangential magnetic field across a resistive sheet, consistent with the modeling of {\em isotropic}, one-layer graphene~\cite{Bludov13}.
By~\eqref{eq:integro-diff}, the sheet is the half line $\Omega=\{ (x,y)\in\mathbb{R}^2: x>0,\, y=0\}$. The solution $u: [0, \infty)\to\C$ expresses the (tangential) $x$-component of the electric field on $\Omega$ (where $\C$ denotes the complex plane); $u^{\rm in}$ is the $x$-component of the {\em incident} electric field on $\Omega$ under transverse-magnetic (TM) polarization; $k$ is the wave number of the ambient medium ($\R k>0$, $\I k\ge 0$); and $\varsigma$ is a dimensionless parameter for the strength of the jump discontinuity across the sheet, with $\varsigma=\omega\mu\sigma/k$ where $\sigma$ is the sheet conductivity and $\mu$ is the magnetic permeability of the ambient medium. Although~\eqref{eq:integro-diff} can be cast into a dimensionless form via the scaling of $x$ with $1/k$, we adhere to~\eqref{eq:integro-diff} for later convenience.

Our goal is to obtain a classical solution to~\eqref{eq:integro-diff}. Our analysis is formal and consists of the following steps. (a) The conversion of~\eqref{eq:integro-diff} into a functional equation with two unknown functions, one of which is the Fourier transform, $\hat u$, of $u$~\cite{WienerHopf1931,PaleyWiener}; (b) the Wiener-Hopf factorization~\cite{Krein62,Masujima-book} to solve the functional equation; (c) the explicit, approximate evaluation, when $u^{\rm in}$ is a plane wave,  of a {\em split function} (defined in Section~\ref{sec:exact}) that is a key ingredient of the factorization process; and (d) the derivation of a one-dimensional (1D) Fourier integral for $u$.
In this procedure, the SPP is defined as a pole contribution to the Fourier integral. The SPP propagates with a wave number $k_{\rm sp}$, $|k_{\rm sp}|\gg |k|$, along the sheet if $\varsigma$ satisfies certain restrictions entirely analogous to those in the three-dimensional (3D) setting~\cite{Margetis16}; see~\eqref{eq:conds-sigma} below.

We relax mathematical rigor, avoiding to precisely describe the space of solutions to~\eqref{eq:integro-diff}. We seek an integrable $u$ on $(0,\infty)$, $u\in L^1(0,\infty)$, assuming that $u^{\rm in}$ is also integrable if the ambient space is lossy ($\I k>0$). The solution for $\I k=0$ is then obtained via analytic continuation.

We also demonstrate that our solution is in good agreement with a numerical computation based on the finite element method. Details of the underlying numerical method lie beyond our present scope; the interested reader may consult~\cite{MML-jcp}.

\subsection{Physical relevance and assumptions}
Physically, the problem at hand is motivated by the need to understand how an SPP can be excited by a TM plane wave incident upon a conducting planar sheet with a defect, i.e., an edge. Our approach yields an integral that connects the relatively rapid spatial variation of the SPP with the behavior of the total diffracted field near the edge. In 2D, the edge acts as an induced point source which radiates a superposition of plane waves with a wide range of wave numbers; depending on $\varsigma$, these plane waves may selectively enable the requisite phase matching for the excitation of the SPP. The Wiener-Hopf method singles out their contribution. In contrast, a plane wave incident on an {\em infinite}, flat conducting sheet {\em cannot} excite an SPP~\cite{Bludov13}.

To solve the Wiener-Hopf functional equation, we use assumptions consistent with low-dimensional metamaterials such as graphene. Specifically, the surface impedance, $\sigma^{-1}$, of the sheet is much larger than the intrinsic impedance of the ambient space~\cite{Bludov13}; and the imaginary part of $\sigma$ is positive. For a slightly lossy ambient medium, these hypotheses amount to having~\cite{Margetis16}
\begin{equation}\label{eq:conds-sigma}
|\varsigma|\ll 1\quad \mbox{and}\quad  \I \varsigma >0~,
\end{equation}
along with $\R\varsigma>0$. By~\eqref{eq:conds-sigma}, we show that the $u$ solving~\eqref{eq:integro-diff} contains a wave, to be defined as the SPP, that has a wavelength much smaller than $(\R k)^{-1}$.

\subsection{On the mathematical approach}
In view of~\eqref{eq:conds-sigma}, one wonders if the Wiener-Hopf method can be replaced by a simpler scheme in seeking a physically transparent approximation for the tangential electric field which would reveal the SPP. Let us heuristically entertain such a scenario of solving~\eqref{eq:integro-diff}. The first condition in~\eqref{eq:conds-sigma} suggests that~\eqref{eq:integro-diff} is amenable to regular perturbations, since $K$ and $u^{\rm in}$ do not depend on $\varsigma$; it is then tempting to solve~\eqref{eq:integro-diff} by successive iterations, starting with the replacement of $u$ by $u^{\rm in}$ under the integral sign. This approach has the flavor of the Neumann expansion for second-kind Fredholm equations~\cite{Masujima-book}; and produces mathematically meaningful terms for $u$ if $u^{\rm in}$ is a plane wave.

However, this regular perturbation fails to {\em directly} produce the SPP.\footnote{A plausible question is whether the SPP can be captured by a re-summation of the ensuing regular-perturbation expansion (in the spirit, e.g., of the Poincar\'e-Lindstedt method for solutions to a class of ordinary differential equations). An answer lies beyond our present scope.} The main reason is that the SPP is expected to have a spatial microstructure, which is proportional to the small parameter, $\varsigma$~\cite{Bludov13,Margetis16}. This feature calls for applying singular perturbations, e.g., a suitable two-scale expansion, for $u$. Alternatively, it is plausible to seek approximations of physical optics for the diffracted field; see Section~\ref{subsec:past} for a discussion of related past works.

The Wiener-Hopf method captures the SPP by a closed-form expression which is amenable to transparent approximations because of~\eqref{eq:conds-sigma}. Thus, this method does not require any unnecessary, a-priori assumption about the structure of this wave. At the risk of redundancy, we repeat that the SPP naturally arises as a pole contribution to a 1D Fourier integral. On the other hand, direct asymptotic methods may prove useful for a {\em finite} strip of graphene~\cite{Myers1965,Satou07}. This problem is not studied in this article.

\subsection{On past works}
\label{subsec:past}
There is a vast body of works in the diffraction of an electromagnetic or acoustic wave
by a half plane. It is impossible to list them exhaustively.

Exact solutions to diffraction problems have been obtained in a limited number of cases. For example, Sommerfeld first solved completely the diffraction of a plane wave from a perfectly conducting half plane via the Fresnel integral~\cite{Sommerfeld1896}; see also~\cite{Carslaw1899}. This technique has been extended to half planes and wedges with the ``impedance boundary condition'', i.e., a linear relation between the tangential electric and magnetic field components, separately on each face of the plane or wedge~\cite{Raman27,Maliuzhinets1958,Williams1959,Kotelnikov13}. Alternate approaches to this type of problem via the Wiener-Hopf functional equation can be found, e.g., in~\cite{Senior1952,Rawlins1975,Hurd1976}.
For comprehensive reviews, the reader may consult~\cite{Weinstein-book,SeniorVolakis-book,LawrieAbrahams07,Castro08,Umul09}.

In principle, the impedance condition expresses a local physical effect different from the condition for a resistive sheet which underlies~\eqref{eq:integro-diff}. The latter condition expresses transmission of the tangential field components across the sheet, thus connecting values of the fields on the two sides of the sheet. However, as pointed in~\cite{Senior1977}, the solution to the problem with impedance conditions can be viewed as a linear superposition of the solution for a resistive sheet and its electromagnetic dual, of the ``magnetically conductive'' sheet, with the magnetic and electric fields interchanged. Thus, the electric current, which is proportional to our $u$, induced on the resistive half plane can be extracted as a particular physical contribution to the current of a suitably chosen problem with impedance boundary conditions. This approach is pursued in~\cite{Senior1979} with recourse to~\cite{Senior1952}.

Our treatment differs from~\cite{Senior1952,Senior1979} in the following three main aspects. First, our formulation spells out the structure of the corresponding boundary value problem for a conducting sheet via integral equation~\eqref{eq:integro-diff}, without direct invocation of an impedance condition. Second, we obtain the solution directly, by means amenable to rigor. In contrast, in~\cite{Senior1952,Senior1979} the author employs a different integral-equation formulation, which amounts to impedance conditions; and extracts a solution for the resistive sheet implicitly. Hence, in~\cite{Senior1952,Senior1979} the analytical linkage of the derived solution to the underlying boundary value problem for a resistive sheet is not evident.

Third, the Wiener-Hopf method here invokes assumptions tailored to the physics of metamaterial sheets such as graphene; cf.~\eqref{eq:conds-sigma}. As a result, a principal ingredient (split function), which depends on the kernel and is needed in the factorization process, is evaluated approximately in terms of a known transcendental function, Euler's dilogarithm. In addition, the SPP emerges as part of our solution. On the other hand, the requisite split function is expressed in terms of an unresolved integral in past works~\cite{Senior1952,Senior1979}; also, in those works the SPP is apparently not part of the solution.

We should mention disparate studies that make use of Keller's geometric theory of diffraction; see, e.g.,~\cite{Keller95,ShiJin08}. In a similar vein, approximations of physical optics focus on the construction of high-frequency solutions to Maxwell's equations away from the edge~\cite{Sehernev76,Bertoni78,Deschamps79,Umul04}. We believe that incorporating the SPP into this physical-optics framework requires some prior knowledge of properties of this wave. Our method amply provides this clue.

\subsection{Article organization and conventions}
\label{subsec:organiz}

The remainder of this article is organized as follows. In Section~\ref{sec:exact}, we derive an exact solution
to~\eqref{eq:integro-diff} via the Wiener-Hopf method. In Section~\ref{sec:approx}, we approximately evaluate a requisite contour integral (split function) by using~\eqref{eq:conds-sigma}, thus simplifying our analytical solution. Section~\ref{sec:comp} focuses on the comparison of our solution to a numerical computation.

Throughout this article, the hat on top of a symbol denotes the Fourier transform of the respective function with respect to $x$. We write $f = \mathcal O(g)$ ($f = o(g)$) to mean that $|f/g|$ is bounded by a nonzero constant (approaches zero) in a prescribed limit; accordingly, $f\sim g$ implies that $f-g=o(g)$. The terms {\em analytic} and {\em holomorphic} function are used interchangeably. The $\pm$ subscript for a function indicates that the function is analytic in the upper ($+$) half, $\mathbb{C}_+$, or lower ($-$) half, $\mathbb{C}_-$, of the complex Fourier domain.

\section{Wiener-Hopf factorization: Exact solution, $u$}
\label{sec:exact}
In this section, we formally seek an exact solution to~\eqref{eq:integro-diff} via the Wiener-Hopf factorization when $u^{\rm in}$ is a plane wave. In Appendix~\ref{app:BVP}, this integral equation is derived from the
requisite boundary value problem. The basic ingredients of the Wiener-Hopf method are reviewed in Appendix~\ref{app:WH}.

\subsection{Main result and definition of SPP}
\label{subsec:main}
The main result of this section is the Fourier integral
\begin{equation}\label{eq:u-FT-exact}
u(x)=\frac{e^{-Q_+(k_\parallel)}}{2\pi i}\int_{-\infty}^\infty \frac{1}{\xi-k_\parallel}\frac{e^{Q_+(\xi)}}{\mathcal P(\xi)}\,e^{i\xi x}\ \db \xi~,\quad x>0~.
\end{equation}
Let us clarify the notation. This formula accounts for $u^{\rm in}(x)=e^{ik_\parallel x}$. We consider $k_\parallel=k\sin\theta$, $0< \theta<\pi/2$, $\I k> 0$, $\R k> 0$; and extend the results to other values of $\theta$ or positive $k$ (as $\I k\downarrow 0$) by analytic continuation. The integration path in the $\xi$-plane lies on the real axis, indented below possible singularities, e.g., $\xi=k_\parallel$, that fall on the real axis from $\mathbb{C}_+$. In~\eqref{eq:u-FT-exact}, the denominator is
\begin{equation}\label{eq:P-denom}
\mathcal P(\xi)=1-i\frac{\varsigma}{k}(k^2-\xi^2)\,\widehat{\mathcal K}(\xi)~,
\end{equation}
where $\widehat{\mathcal K}(\xi)$ is the Fourier transform of the kernel, viz., (see Appendix~\ref{app:BVP})
\begin{equation}\label{eq:K-FT-top}
\widehat{\mathcal K}(\xi)=\int_{-\infty}^{\infty}\mathcal K(x)\,e^{-i\xi x}\,\db x=\frac{i}{2\sqrt{k^2-\xi^2}}~,\quad \I\sqrt{k^2-\xi^2}>0~,
\end{equation}
in the top Riemann sheet for $\sqrt{k^2-\xi^2}$ ($\xi\in\mathbb{C}$). The function $Q_+(\xi)$ entering~\eqref{eq:u-FT-exact} is analytic in the upper half, $\mathbb{C}_+$, of the $\xi$-plane; and is defined by the (absolutely) convergent integral
\begin{equation}\label{eq:Q+_form}
Q_+(\xi)=\frac{\xi}{\pi i}\int_{0}^\infty \frac{Q(\zeta)}{\zeta^2-\xi^2}\ \db\zeta~,\quad \xi\in\mathbb{C}_+~;\quad Q(\zeta):=\ln\mathcal P(\zeta)~.
\end{equation}
This $Q_+$ is a `$+$' split function.
Note that $\mathcal P(\zeta)$ and $Q(\zeta)$ are even in the top Riemann sheet. We have assumed that  (see Definition~\ref{def:index} in Appendix~\ref{app:WH})
\begin{equation*}
{\rm ind}(\mathcal P)=0~,
\end{equation*}
which means that $Q(\zeta)=\ln\mathcal P(\zeta)$ does {\em not} pick up any phase change of $\mathcal P(\zeta)$ as $\zeta$ moves from $-\infty$ to $+ \infty$ in~\eqref{eq:Q+_form}. The $Q_+(\xi)$ is evaluated in Section~\ref{sec:approx}.

By resorting to~\eqref{eq:u-FT-exact}, we can now define the SPP more precisely.
\begin{definition}[SPP]\label{def:SP}
By~\eqref{eq:u-FT-exact}, the SPP is the residue at the (simple) pole, $\xi=k_{\rm sp}$, of $\mathcal P(\xi)^{-1}$ in the upper half of the $\xi$-top Riemann sheet. If the pole contributes, the SPP is described as part of $u$ by the formula
\begin{equation}\label{eq:u-sp}
u^{\rm sp}(x)=e^{ik_{\rm sp}x}\,\frac{e^{-Q_+(k_\parallel)+Q_+(k_{\rm sp})}}{k_{\rm sp}-k_\parallel}\frac{1}{\mathcal P'(k_{\rm sp})}~,\quad x>0~,
\end{equation}
where $\mathcal P(k_{\rm sp})=0$ and $\I k_{\rm sp}>0$ with
$\I\sqrt{k^2-\xi^2}\big|_{\xi=k_{\rm sp}} >0$. (Here, $\mathcal P'(\xi)$ denotes the first derivative of $\mathcal P(\xi)$.)
\end{definition}
It can be readily verified that, for sufficiently small $\I k$, this $k_{\rm sp}$ is present in the top Riemann sheet with $|k_{\rm sp}|\gg |k|$ provided~\eqref{eq:conds-sigma} hold. By~\eqref{eq:P-denom}, we find
\begin{equation}\label{eq:ksp}
k_{\rm sp}=i\frac{2k}{\varsigma}\sqrt{1-\varsigma^2}\sim i\frac{2k}{\varsigma}\quad \mbox{if}\ |\varsigma|\ll 1~.
\end{equation}
This $k_{\rm sp}$ is identical with the SPP wave number in the 3D setting~\cite{Bludov13,Margetis16}.
\begin{remark}\label{rmk:diffract}
It is of interest to comment on the contribution of the pole $\xi=k_\parallel$ to integral~\eqref{eq:u-FT-exact}. The respective residue is the ``direct'' field
\begin{equation}\label{eq:u-dir}
u^{\rm dir}(x)=\frac{e^{ik_\parallel x}}{\mathcal P(k_\parallel)}~,\quad x>0~,
\end{equation}
which is the sum of the incident field, $u^{\rm in}$, and its direct reflection from an infinite conducting plane. Thus, the remaining, diffracted field for $x>0$ is
\begin{align}\label{eq:diffr-u}
u^{\rm df}(x)&=u(x)-u^{\rm dir}(x)\notag\\
&=\frac{e^{-Q_+(k_\parallel)}}{2\pi i}\int_{-\infty}^\infty \frac{1}{\xi-k_\parallel}\Biggl[\frac{e^{Q_+(\xi)}}{\mathcal P(\xi)}-\frac{e^{Q_+(k_\parallel)}}{\mathcal P(k_\parallel)}\Biggr]\,e^{i\xi x}\,\db \xi~,
\end{align}
which contains the SPP by Definition~\ref{def:SP}. In Section~\ref{sec:approx}, we show that the SPP is the dominant contribution to $u^{\rm df}$ under~\eqref{eq:conds-sigma} for a certain range of distances, $x$, from the edge.
\end{remark}

\subsection{Derivation of exact solution}
\label{subsec:derive}

This section is devoted to the derivation of~\eqref{eq:u-FT-exact} by virtue of~\eqref{eq:integro-diff}. We assume that $\I k> 0$ and $u$ is integrable on $(0,\infty)$. By setting $u(x)\equiv 0$ and $u^{\rm in}(x)\equiv 0$ for $x<0$, we write~\eqref{eq:integro-diff} as
\begin{equation}\label{eq:ext}
u(x)=u^{\rm in}(x)+g(x)+\frac{i\varsigma}{k}\biggl(\frac{\db^2}{\db x^2}+k^2\biggr)
\int_{-\infty}^\infty \mathcal K(x-x')\,u(x')\,\db x'~,\quad x\in \mathbb{R}~,
\end{equation}
where
\begin{equation*}
g(x):=\left\{\begin{array}{lr} {\displaystyle -i\frac{\varsigma}{k}\biggl(\frac{\db^2}{\db x^2}+k^2\biggr)\int_0^{\infty}\mathcal K(x-x')\,u(x')\,\db x'}~,& x<0~,\\
0~,& x>0~.\end{array}\right.
\end{equation*}
The application of the Fourier transform to~\eqref{eq:ext} yields the functional equation
\begin{equation}\label{eq:func-eq}
\mathcal P(\xi)\widehat{u}(\xi)=\widehat{u}^{\rm in}(\xi)+\widehat{g}(\xi)~,\quad \xi\in\mathbb{R}~;
\end{equation}
cf.~\eqref{eq:WE-u-FT} of Appendix~\ref{app:WH}. Clearly, $\widehat{u}(\xi)$ is analytic in $\mathbb{C}_-$ and $\widehat{g}(\xi)$ is analytic in $\mathbb{C}_+$. The $\mathcal P(\xi)$ is dependent on the kernel and is defined in~\eqref{eq:P-denom}.

The task is to separate the terms in~\eqref{eq:func-eq} into two parts; one part should be holomorphic in $\mathbb{C}_+$ and another in $\mathbb{C}_-$ (see Appendix~\ref{app:WH}). First, we need to determine split functions $Q_s(\xi)$, holomorphic in $\mathbb{C}_s$ ($s=\pm$), such that
\begin{equation}\label{eq:P-factoriz}
Q(\xi)=Q_+(\xi)+Q_-(\xi)~,\quad \xi\in\mathbb{R}~;\quad \mathcal P(\xi)=e^{Q(\xi)}~.
\end{equation}
Since $Q(\xi)$ is analytic in a neighborhood of the real axis and an even function, with ${\rm ind}(\mathcal P)=0$, application of the Cauchy integral formula yields~\cite{Krein62}
\begin{align}\label{eq:Qs-Cauchy}
Q_s(\xi)&=s\frac{1}{2\pi i}\int_{-\infty}^\infty \frac{Q(\zeta)}{\zeta-\xi}\ \db\zeta\nonumber\\
&=s\frac{\xi}{\pi i}\int_{0}^\infty \frac{Q(\zeta)}{\zeta^2-\xi^2}\ \db\zeta~;\quad s\I \xi>0\quad  (s=\pm)~.
\end{align}

It follows that~\eqref{eq:func-eq} is recast to
\begin{equation}\label{eq:func-eqII}
e^{Q_-(\xi)}\widehat{u}(\xi)=e^{-Q_+(\xi)}\widehat{u}^{\rm in}(\xi)+e^{-Q_+(\xi)}\widehat{g}(\xi)~,\quad \xi\in\mathbb{R}~,
\end{equation}
where the left-hand side is analytic in $\mathbb{C}_-$ and the last term on the right-hand side is analytic in $\mathbb{C}_+$. To split the remaining term, note that
\begin{equation*}
\widehat{u}^{\rm in}(\xi)=\frac{-i}{\xi-k_{\parallel}}~,\quad \I k_\parallel >0~.
\end{equation*}
Thus, by setting
\begin{equation*}
e^{-Q_+(\xi)}\widehat{u}^{\rm in}(\xi)=\Lambda_+(\xi)+\Lambda_-(\xi)~,
\end{equation*}
we obtain the requisite split functions by inspection, viz.,
\begin{equation}\label{eq:Lambda+-}
\Lambda_+(\xi)=\bigl[e^{-Q_+(\xi)}-e^{-Q_+(k_\parallel)}\bigr]\widehat{u}^{\rm in}(\xi)~,\quad \Lambda_-(\xi)=e^{-Q_+(k_\parallel)}\widehat{u}^{\rm in}(\xi)~.
\end{equation}
Consequently, \eqref{eq:func-eqII} becomes
\begin{equation}\label{eq:func-eqIII}
e^{Q_-(\xi)}\widehat{u}(\xi)-\Lambda_-(\xi)=\Lambda_+(\xi)+e^{-Q_+(\xi)}\widehat{g}(\xi)~,\quad \xi\in\mathbb{R}~,
\end{equation}
where the `$+$' and `$-$' parts are completely separated. Thus, we infer
that the two sides of this equation together define an {\em entire} function, $E(\xi)$, in the complex $\xi$-plane. In particular, we have
\begin{equation}\label{eq:entire-E}
E(\xi)=\left\{\begin{array}{lr} e^{Q_-(\xi)}\widehat{u}(\xi)-\Lambda_-(\xi)~,& \I\xi<0~,\\
\Lambda_+(\xi)+e^{-Q_+(\xi)}\widehat{g}(\xi)~,& \I\xi>0~.\end{array}
\right.
\end{equation}
Once this $E(\xi)$ is determined, $\widehat{u}(\xi)$ is found via~\eqref{eq:entire-E}, viz.,
\begin{equation}\label{eq:u-by-E}
\widehat{u}(\xi)=e^{-Q_-(\xi)}[\Lambda_-(\xi)+E(\xi)]~.
\end{equation}

Next, we show that $E(\xi)\equiv 0$ via the asymptotic behavior of~\eqref{eq:entire-E} as $\xi\to\infty$. Since $u(x)$ is integrable on $(0,\infty)$ and vanishes identically on $(-\infty,0)$, we assert that $\widehat{u}(\xi)\to 0$ as $\xi\to\infty$ in $\mathbb{C}_-$. In a similar vein, $\widehat{g}(\xi)\to 0$ as $\xi\to\infty$ in $\mathbb{C}_+$. Hence, in order to extract $E(\xi)$, we need to know how $Q_\pm(\xi)$ behave as $\xi\to\infty$ in $\mathbb{C}_\pm$, respectively.

Let us focus on the $Q_+(\xi)$ by virtue of~\eqref{eq:Qs-Cauchy}. Notice that
\begin{equation*}
Q(\zeta)=\ln\biggl(\frac{i\varsigma}{2k}\zeta\biggr)+\widetilde Q(\zeta)~;\quad \widetilde Q(\zeta)=\mathcal O(1/\zeta)\quad\mbox{as}\ \zeta\to +\infty~.
\end{equation*}
Hence, if we naively approximate the denominator in~\eqref{eq:Qs-Cauchy} according to $\zeta^2-\xi^2\approx -\xi^2$ for large $\xi$, the resulting integral diverges. Therefore, to extract the leading-order term for $Q_+(\xi)$ as $\xi\to\infty$, we write
\begin{equation*}
Q_+(\xi)=\frac{\xi}{i\pi}\int_0^\infty
\frac{\ln\bigl(\frac{i\varsigma}{2k}\zeta\bigr)}{\zeta^2-\xi^2}\,\db\zeta+\frac{1}{i\pi}\int_0^\infty \frac{\widetilde{Q}(\zeta)}{(\zeta/\xi)^2-1}\,\db(\zeta/\xi)~,\quad \I\xi>0~.
\end{equation*}
The second term approaches zero as $\xi\to\infty$; and the first term is an elementary integral which is computed exactly by closing the path in the $\zeta$-plane through the negative real axis and a large
semicircle in $\mathbb{C}_+$. Hence, we obtain
\begin{equation}\label{eq:Q+-asympt}
Q_+(\xi)=\frac{1}{2}\ln\biggl(\varsigma\,\frac{\xi}{2k}\biggr)+o(1)\quad\mbox{as}\ \xi\to\infty\ \mbox{in}\ \mathbb{C}_+~.
\end{equation}
By symmetry, we have $Q_-(\xi)=(1/2)\ln[-\varsigma\xi/(2k)]+o(1)$ as $\xi\to\infty$ in $\mathbb{C}_-$.

Equations~\eqref{eq:Lambda+-} then entail $\Lambda_\pm(\xi)\to 0$ as $\xi\to\infty$ in $\mathbb{C}_\pm$. Thus, by~\eqref{eq:entire-E} we assert that $E(\xi)\to 0$ as $\xi\to\infty$ in $\mathbb{C}$. By Liouville's theorem, $E(\xi)$ must be a constant everywhere, which has to be zero; $E(\xi)\equiv 0$. Consequently,
formula~\eqref{eq:u-FT-exact} results from Fourier-inversion of~\eqref{eq:u-by-E} by use of~\eqref{eq:P-factoriz} and~\eqref{eq:Lambda+-}.

\section{Approximate split function and solution}
\label{sec:approx}
In this section, we approximately evaluate $Q_+(\xi)$, which enters formula~\eqref{eq:u-FT-exact} for $u(x)$. To this end, we enforce conditions~\eqref{eq:conds-sigma}. Accordingly, we obtain an approximation for $u(x)$ by manipulating its 1D Fourier integral.

The main result of this section is the formula
\begin{equation}\label{eq:Q+_approx-fin}
Q_+(\xi)\sim \frac{1}{2\pi i}\frac{\varsigma\xi}{k}[\mathcal F_1(\xi/k)+\mathcal F_2(\varsigma\xi/k)]~,\quad |\varsigma|\ll 1~,
\end{equation}
where the functions $\mathcal F_j$ ($j=1,2$) are described in~\eqref{eq:F1} and~\eqref{eq:F2} below. Notably, this approximation for $Q_+(\xi)$ exhibits two distinct (fast and slow) scales in the complex $\xi$-plane.

\subsection{Approximation for $Q_+(\xi)$}
\label{subsec:Q+_approx}
Equation~\eqref{eq:Qs-Cauchy}, with $s=+$, is conveniently recast into the expression
\begin{align*}
Q_+(\xi)=&\frac{\xi}{i\pi}\Biggl\{ \int_0^\infty
\frac{{\displaystyle \ln\bigl(1+\frac{\varsigma}{2k}\sqrt{k^2-\zeta^2}\bigr)-\ln\bigl(1+\frac{\varsigma}{2k}i\zeta\bigr)}}{\zeta^2-\xi^2}\ \db\zeta \notag\\
&\quad +\int_0^\infty \frac{{\displaystyle \ln\bigl(1+\frac{\varsigma}{2k}i\zeta\bigr)}}{\zeta^2-\xi^2}\ \db\zeta\Biggr\}~.
\end{align*}
The first integral is amenable to asymptotics if $|\varsigma|\ll 1$. A key observation is that the integrand of this integral has a negligible contribution if $\zeta\gg |k|$. By expanding out its logarithms, we write
\begin{equation}\label{eq:Q+_I12}
Q_+(\xi) \sim \frac{1}{2\pi i}\frac{\varsigma \xi}{k} \,\biggl\{\mathcal I_1(\xi)+\mathcal I_2(\xi)\biggr\}~,\quad |\varsigma|\ll 1~,
\end{equation}
where
\begin{subequations}
\begin{align}
\mathcal I_1(\xi)&=\int_0^\infty \frac{\sqrt{k^2-\zeta^2}-i\zeta}{\zeta^2-\xi^2}\ \db\zeta= \int_0^{\infty e^{-i\arg k}} \frac{\sqrt{1-\tau^2}-i\tau}{\tau^2-(\xi/k)^2}\ \db \tau~,\label{eq:I1}\\
\mathcal I_2(\xi)&=\biggl(\frac{\varsigma}{2k}\biggr)^{-1}\int_0^\infty \frac{{\displaystyle \ln\bigl(1+\frac{i\varsigma}{2k}\zeta}\bigr)}{\zeta^2-\xi^2}\ \db\zeta\label{eq:I2}\\
&=\int_0^{\infty e^{i\arg(\varsigma/k)}}\frac{\ln(1+i\tau)}{\tau^2-(\xi\varsigma/(2k))^2}\,\db\tau~.\notag
\end{align}
\end{subequations}
In the above, we applied transformations $\zeta\mapsto \tau$ in order to spell out the dependence of the integrals on the physical parameters $k$ and $\varsigma$. In~\eqref{eq:I1}, $\tau=\zeta/k$; and in~\eqref{eq:I2}, $\tau=(\varsigma/(2k))\zeta$.
Our next task is to compute $\mathcal I_1$ and $\mathcal I_2$ exactly.

\subsubsection{Integral $\mathcal I_1(\xi)$}
\label{sssec:I1}
To simplify the derivation, first consider $\R k=0$ with $\I k>0$. We can continue analytically the result to the desired value of $k$ in the end. By the change of variable $\zeta\mapsto \nu$ under $\zeta=ik\sinh(\ln \nu/2)$ and the substitution $\xi=-ik\sinh \varpi$, integral~\eqref{eq:I1} becomes
\begin{subequations}\label{eq:I1-fin}
\begin{align}\label{eq:I1-a}
\mathcal I_1(\xi)&=i\int_0^1 \frac{1+\nu}{\nu^2-2(\cosh(2\varpi))\nu+1}\ \db\nu	\nonumber\\
&= \frac{ie^{\varpi}}{2\sinh\varpi}\ln(1-e^{-2\varpi})-\frac{ie^{-\varpi}}{2\sinh\varpi}\ln(1-e^{2\varpi})\notag\\
&= \mathcal F_1(\xi/k)~,
\end{align}
where
\begin{align}\label{eq:F1}
\mathcal F_1(w)&=\frac{\sqrt{1-w^2}}{w}\,\ln\bigl(w+i\sqrt{1-w^2}\bigr)+i\ln(2w)~,
\end{align}
\end{subequations}
which was computed by elementary methods.
Note in passing that this result reads
$\mathcal I_1(\xi)=i(1-\varphi^2)^{-1/2}\varphi\arcsin\varphi+i\ln(2\sqrt{1-\varphi^2})$ where
$\varphi(\xi)=\sqrt{1-(\xi/k)^2}$. The reader may observe that this function does not have any singularity, e.g., a branch point, at $\varphi=0$. Thus, $\mathcal I_1(\xi)$ does not have any singularity at $\xi=\pm k$. Evidently, $\mathcal I_1(\xi)$ has a branch point at $\xi=0$.

\subsubsection{Integral $\mathcal I_2(\xi)$}
\label{sssec:I2}
We now turn our attention to~\eqref{eq:I2}. Notice that
\begin{equation}\label{eq:I2-F}
\mathcal I_2(\xi)=\frac{k}{\varsigma\xi}\lim_{M\to +\infty}\bigl[F(\varsigma\xi/k, M)-F(-\varsigma\xi/k,M)\bigr]~,	
\end{equation}
where
\begin{equation}
F(w,M):=\int_0^M \frac{\ln(1+i\zeta/2)}{\zeta-w}\ \db\zeta~.
\end{equation}
This $F$ can be evaluated in terms of Euler's dilogarithm, $L_2$, defined by~\cite{BatemanI}
\begin{equation*}
L_2(-z):=-\int_0^z \frac{\ln(1+u)}{u}\ \db u~.
\end{equation*}
We directly compute
\begin{equation*}
F(w,M)=\ln\biggl(1+\frac{i}{2}w\biggr)\,
\ln\biggl(\frac{M-w}{-w}\biggr)-L_2(-A(w,M))+L_2(-A(w,0))~,
\end{equation*}
where
\begin{equation*}
A(w,\ell):=\frac{i}{2}(\ell-w)
\biggl(1+\frac{i}{2}w\biggr)^{-1}~.
\end{equation*}

In the limit $M\to\infty$, we need to invoke the asymptotic expansion of $L_2(z)$ for large $|z|$~\cite{BatemanI}. Consequently, we are able to verify that the right-hand side of~\eqref{eq:I2-F} indeed yields a finite limit for fixed $\varsigma \xi/k$ as $M\to\infty$. After some algebra, the details of which we leave to the reader, we obtain
\begin{subequations}\label{eq:I2-fin}
\begin{equation}\label{eq:I2-a}
\mathcal I_2(\xi)=\mathcal F_2(\varsigma \xi/k)~,
\end{equation}
where
\begin{align}\label{eq:F2}
\mathcal F_2(w)&=\frac{1}{w}\Biggl\{
\ln\biggl(\frac{1-iw/2}{1+iw/2}\biggr)\,\ln\biggl[\frac{iw}{2}(1+w^2/4)^{-1/2}\biggr]+i\pi \ln(1+iw/2)\notag\\
&\qquad +L_2(-A(w,0))-L_2(-A(-w,0))\}~.
\end{align}
\end{subequations}
It can be readily checked that this $\mathcal F_2(w)$, for $w=\varsigma\xi/k$, does not have any singularity, e.g., a branch point, at $\xi=i\,2k/\varsigma\sim k_{\rm sp}$ ($w=2i$). In fact, the logarithmic singularity at $w=2i$ that is present in the first line of~\eqref{eq:F2} turns out to be canceled by the asymptotic behavior of Euler's dilogarithm as $A(w,0)\to \infty$.

Equation~\eqref{eq:Q+_I12} along with~\eqref{eq:I1-fin} and~\eqref{eq:I2-fin} yield the desired formula for $Q_+(\xi)$; cf.~\eqref{eq:Q+_approx-fin}.

\subsubsection{Particular values of $Q_+(\xi)$}
\label{sssec:special_v}
It is of interest to compute two values of $Q_+(\xi)$, which are needed for the description of the solution, $u$; see Section~\ref{sssec:diffracted}. We assume that $k_\parallel=\mathcal O(1)$ as
 $\varsigma\to 0$. By~\eqref{eq:Q+_approx-fin}, we find
\begin{subequations}\label{eq:Q+_spec}
\begin{align}
Q_+(k_\parallel)&= \frac{1}{2\pi}\frac{k_\parallel}{k}\varsigma[\ln(4/\varsigma)+1]+\frac{\varsigma}{2\pi}\biggl[\frac{\pi}{2}-\arcsin\biggl(\frac{k_\parallel}{k}\biggr)\biggr]\sqrt{1-(k_\parallel/k)^2}\notag\\
&\quad +\mathcal O(\varsigma^2 \ln\varsigma)~,\label{eq:Q+_kpar}\\
Q_+(k_{\rm sp})&= \frac{1}{4\pi i}\left[ -\ln^2 2+2\pi i \ln 2-\frac{\pi^2}{3}-2L_2(1/2)\right]+\mathcal O(\varsigma^2 \ln\varsigma)~,\label{eq:Q+_ksp}
\end{align}
\end{subequations}
for $|\varsigma|\ll 1$.
Note that~\eqref{eq:Q+_kpar} includes the case with small or vanishing $k_\parallel/k$, in which the second term of the asymptotic expansion may dominate; and the case $k_\parallel=k$ in which the first term prevails. Hence, we assert that
\begin{align*}
e^{-Q_+(k_\parallel)}&=1-\frac{1}{2\pi}\frac{k_\parallel}{k}\varsigma[\ln(4/\varsigma)+1]-\frac{\varsigma}{2\pi}\biggl[\frac{\pi}{2}-\arcsin\biggl(\frac{k_\parallel}{k}\biggr)\biggr]\sqrt{1-(k_\parallel/k)^2}\notag\\
&\quad +\mathcal O(\varsigma^2 \ln\varsigma)~.
\end{align*}

\subsection{Approximations for $u$}
\label{subsec:approx_soln}
Next, we heuristically discuss the relative importance of the SPP in the diffracted field, formula~\eqref{eq:diffr-u}, in light of approximation~\eqref{eq:Q+_I12}.
Furthermore, we describe the total field, $u(x)$, if $|k_{\rm sp}x|\ll 1$, i.e., near the edge of the conducting sheet.

\subsubsection{On the diffracted field, $u^{\rm df}$}
\label{sssec:diffracted}
By formula~\eqref{eq:diffr-u}, we write
\begin{equation*}
u^{\rm df}(x)=u^{\rm sp}(x)+u^{\rm rad}(x)~,\qquad x>0~,
\end{equation*}
where, in view of Definition~\ref{def:SP}, we introduce the radiation field
\begin{align}
u^{\rm rad}(x)&=\frac{e^{-Q_+(k_\parallel)}}{2\pi i}\int_{-\infty}^\infty \frac{1}{\mathcal P(\xi)}\biggl\{\biggl[e^{Q_+(\xi)}-\frac{\mathcal P(\xi)}{\mathcal P(k_\parallel)}e^{Q_+(k_\parallel)}\biggr]\frac{1}{\xi-k_\parallel}\notag\\
&\hphantom{=	u^{\rm sp}(x)+\frac{e^{-Q_+(k_\parallel)}}{2\pi i}}-\frac{e^{Q_+(k_{\rm sp})}}{k_{\rm sp}-k_\parallel}\frac{\mathcal P(\xi)}{\mathcal P'(k_{\rm sp})\,(\xi-k_{\rm sp})}\biggr\}\,e^{i\xi x}\ \db x~,\label{eq:u-rad}
\end{align}
along with the SPP contribution, viz.,
\begin{align}\label{eq:SP-form}
u^{\rm sp}(x)&=\biggl(\frac{2k}{\varsigma}\biggr)^2\frac{1}{k_{\rm sp}(k_{\rm sp}-k_\parallel)}e^{-Q_+(k_\parallel)+Q_+(k_{\rm sp})}e^{ik_{\rm sp}x}\notag\\
&\sim -e^{-Q_+(k_\parallel)+Q_+(k_{\rm sp})}e^{ik_{\rm sp}x}~;
\end{align}
recall formulas~\eqref{eq:Q+_approx-fin} and~\eqref{eq:Q+_spec} for $Q_+(\xi)$, and
$k_{\rm sp}=i(2k/\varsigma)\sqrt{1-\varsigma^2}\sim i2k/\varsigma$.

Let us now focus on~\eqref{eq:u-rad}. The only singularity of the integrand in the upper half of the $\xi$-plane, $\mathbb{C}_+$, is the branch point at $\xi=k$, because of $\mathcal P(\xi)$. Although we have not been able to express the requisite integral in terms of known transcendental functions for all $x$, it is possible to simplify it in certain regimes of $kx$.

By deforming the integration path in $\mathbb{C}_+$ for $x>0$, and wrapping the path around the branch cut that emanates from $\xi=k$, we directly obtain
\begin{align}\label{eq:u-rad-bc}
u^{\rm rad}(x)\sim&\frac{\varsigma}{2\pi}e^{ik x-i\pi/4}	\int_0^\infty \frac{e^{Q_+(k(1+i\tau))}}{1-k_\parallel/k +i\tau}\notag\\
&\qquad \times \frac{\sqrt{\tau}\sqrt{2+i\tau}}{1+(i\varsigma^2/4)\tau(2+i\tau)}\,e^{-(kx)\tau}\ \db\tau~.
\end{align}
This formula is amenable to comparisons to $u^{\rm sp}(x)$.

First, consider $kx=\mathcal O(1)$. By treating $\tau$ as an $\mathcal O(1)$ variable, we can use $Q_+(k(1+i\tau))=\mathcal O(\varsigma\ln\varsigma)$ as $\varsigma\to 0$; thus, we end up with the approximation
\begin{equation}\label{eq:u-rad-kx}
u^{\rm rad}(x) \sim
\frac{\varsigma}{2\pi}	e^{ikx-i\pi/4}\,h(kx)~,
\end{equation}
where
\begin{equation*}
h(z):=\int_0^\infty \frac{\sqrt{\tau}\sqrt{2+i\tau}}{i\tau+1-k_\parallel/k}	\,e^{-z\tau}\ \db\tau~.
\end{equation*}
Clearly, by~\eqref{eq:SP-form}, the SPP dominates if
\begin{equation}
e^{[\R(2k/\varsigma)-\I k]x}\ll |\varsigma|^{-1}~;\qquad kx=\mathcal O(1)~.
\end{equation}

We can still use and simplify~\eqref{eq:u-rad-kx} when $|kx|\gg 1$, uniformly in the parameter $1-k_\parallel/k$. For this purpose, consider the identity~\cite{BatemanI}
\begin{equation*}
\frac{\db}{\db z}\bigl[e^{iz(1-k_\parallel/k)}h(z)\bigr]=e^{-i3\pi/4}\frac{\pi}{2}\frac{e^{-i(k_\parallel/k)z}}{z}H_1^{(1)}(z)~,
\end{equation*}
where $H_1^{(1)}(z)$ is the first-kind Hankel function
of order one. Integration of the last equation entails
\begin{equation}
u^{\rm rad}(x)\sim \frac{\varsigma}{4}\,e^{ik_\parallel x}\int_{kx}^\infty \frac{H_1^{(1)}(t)}{t}\,e^{-i(k_\parallel/k)t}\ \db t~.
\end{equation}
Hence, if $|kx|\gg 1$, we may use the asymptotic expansion of $H_1^{(1)}(t)$ for large $|t|$~\cite{BatemanII}. After some algebra, we find
\begin{align}\label{eq:u-rad-larg-kx}
u^{\rm rad}(x)&\sim \frac{\varsigma}{\sqrt{2\pi}}\,e^{ik_\parallel x-i3\pi/4}	 \biggl(\frac{k-k_\parallel}{k}\biggr)^{1/2}\left[\frac{e^{i(k-k_\parallel)x}}{\sqrt{(k-k_\parallel)x}}+\mathcal F((k-k_\parallel)x)\right]~,
\end{align}
where
\begin{equation*}
\mathcal F(z):=i\sqrt{2\pi}\bigl[2^{-1/2}e^{i\pi/4}	-C(z)-iS(z)\bigr]~,
\end{equation*}
and $C(z)$ and $S(z)$ are Fresnel integrals, defined by~\cite{BatemanI}
\begin{equation*}
C(z)=\frac{1}{\sqrt{2\pi}}\int_0^z\frac{\cos t}{\sqrt{t}}\ \db t~,\quad S(z)=\frac{1}{\sqrt{2\pi}}\int_0^z\frac{\sin t}{\sqrt{t}}\ \db t~.
\end{equation*}
Note the expansion
\begin{equation*}
\mathcal F(z)\sim e^{iz}\biggl(-\frac{1}{\sqrt{z}}+\frac{i}{2z^{3/2}}\biggr)\quad \mbox{as}\ |z|\to\infty~,	\end{equation*}
which is useful if $|(k-k_\parallel)x|\gg 1$, along with $|kx|\gg 1$.

Next, consider $k_{\rm sp}x=\mathcal O(1)$, i.e., $kx=\mathcal O(\varsigma)$. The major contribution to integration in~\eqref{eq:u-rad-bc} comes from values of $\tau$ such that $\varsigma\tau=\mathcal O(1)$. By~\eqref{eq:Q+_approx-fin},
\begin{equation*}
Q_+(k(1+i\tau))\sim (2\pi i)^{-1}\varsigma(1+i\tau)\mathcal F_2(\varsigma(1+i\tau))~.	
\end{equation*}
Indeed, this formula suggests that $\varsigma\tau$ is the natural, $\mathcal O(1)$ variable in this regime; then, $Q_+(k(1+i\tau))=\mathcal O(1)$. The scaling of $\tau$ by $\varsigma^{-1}$ implies that
\begin{equation}
e^{-ikx}\,u^{\rm rad}(x)=\mathcal O(1)\quad\mbox{if}\ k_{\rm sp}x=\mathcal O(1)~.
\end{equation}
Hence, in this regime, $|u^{\rm rad}|$ can be comparable to or smaller than $|u^{\rm sp}|$.

Now let us assume that $1\ll |k_{\rm sp}x|< \mathcal O(|\varsigma|^{-1})$, the major contribution to integration for
$u^{\rm rad}(x)$ comes from $\varsigma\tau=\mathcal O((k_{\rm sp}x)^{-1})$. It follows that
\begin{equation}
e^{-ikx}\,u^{\rm rad}(x)=\mathcal O((k_{\rm sp}x)^{-1})~.	
\end{equation}
The SPP dominates if
\begin{equation*}
e^{[\R(2k/\varsigma)-\I k]x}\ll |2kx|\,|\varsigma|^{-1}~.	\end{equation*}

\subsubsection{Near-edge field, $|k_{\rm sp}x|\ll 1$}
\label{sssec:near-f}
In this case, we need to examine the total field, $u(x)$. As $x\downarrow 0$, this solution is expected to be singular, i.e., the first derivative of $u(x)$ blows up.

First, by~\eqref{eq:u-FT-exact} we notice that $\lim_{x\downarrow 0}u(x)=0$.
Accordingly, in order to derive the asymptotic behavior of $u(x)$ as $x\downarrow 0$, it suffices to compute $\db u/\db x$ for small $x$ and then integrate the result.
Bear in mind that
\begin{equation}\label{eq:dudx}
\frac{\db u}{\db x}=\frac{e^{-Q_+(k_\parallel)}}{2\pi}\int_{-\infty}^\infty \frac{\xi}{1+\frac{\varsigma}{2k}\sqrt{k^2-\xi^2}}\,\frac{e^{Q_+(\xi)}}{\xi-k_\parallel}\,e^{i\xi x}\ \db\xi~.
\end{equation}

Second, as $x\downarrow 0$, in~\eqref{eq:dudx} the major contribution to integration arises from large
$\xi$,
$|\xi|\gg |k_{\rm sp}|$, in $\mathbb{C}_+$. Thus, by properly deforming the path in
$\mathbb{C}_+$, we simplify $Q_+(\xi)$ according to
\begin{equation*}
Q_+(\xi)\sim \frac{1}{2\pi i}\frac{\varsigma\xi}{k}\,F_2(\varsigma\xi/k)= \frac{1}{2}\ln\biggl(\frac{\varsigma\xi}{2k}\biggr)+\mathcal O(1/\xi)~,
\end{equation*}
in agreement with~\eqref{eq:Q+-asympt}. Consequently, we compute
\begin{align*}
\frac{\db u}{\db x}&\sim \frac{e^{-Q_+(k_\parallel)}}{2\pi}\int_{-\infty}^\infty \frac{\xi}{\frac{\varsigma}{2k}i\xi}\,\frac{e^{(1/2)\ln[\varsigma\xi/(2k)]}}{\xi}\,e^{i\xi x}\,\db\xi\notag\\
&\sim e^{-i\pi/4-Q_+(k_\parallel)}\sqrt{\frac{2}{\pi}}\,\frac{k}{\sqrt{\varsigma kx}}\quad \mbox{as}\ x\downarrow 0~.
\end{align*}
Hence, we obtain
\begin{equation}\label{eq:u-asympt}
u(x)=\int_0^x \frac{\db u(z)}{\db z}\ \db z\sim 2e^{-i\pi/4-Q_+(k_\parallel)}\sqrt{\frac{2kx}{\pi\varsigma}}\quad \mbox{as}\ x\downarrow 0~.
\end{equation}
Notice that the singularity is strengthened by the factor $\varsigma^{-1/2}$ and $e^{-Q_+(k_\parallel)}\sim 1$  ($|\varsigma|\ll 1$).

\section{A numerical computation}
\label{sec:comp}

In this section, we compare our analytical findings against a direct
numerical computation with curl-conforming finite elements. For details of
the underlying numerical method, we refer the reader to \cite{MML-jcp}. For the numerical 
computation, we set $k=1$ and fix the remaining parameters by choosing the values
\begin{align*}
  \varsigma = 0.002 + 0.2\,i~,\quad
  \theta=\pi/9~.
\end{align*}

\subsection{On the numerical computation of $Q_+$}
\label{subsec:num-Q+}

\begin{table}[!tbp]
  \centering
  \begin{tabular}{l c c}
    \toprule
    & $Q_+(k_\parallel)$ & $Q_+(k_{\rm sp})$\\[0.3em]
    Approx.~\eqref{eq:Q+_spec}       & 0.017793 + 0.079874\,i & 0.34657 + 0.39270\,i \\
    Approx.~\eqref{eq:Q+_approx-fin} & 0.017507 + 0.079995\,i & 0.34793 + 0.39902\,i \\
    Definition~\eqref{eq:Q+_form}    & 0.019864 + 0.079563\,i & 0.34813 + 0.40071\,i \\
    \bottomrule
  \end{tabular}
  \caption{Comparison of approximations for $Q_+(\xi)$, particularly formulas
    \eqref{eq:Q+_spec} and \eqref{eq:Q+_approx-fin}, against the defining, exact
    integral~\eqref{eq:Q+_form}.}
  \label{tab:qplus}
\end{table}

Next, we indicate the accuracy of our approximate formulas for $Q_+(\xi)$ by selectively computing this function for $\xi=k_\parallel$ and $\xi=k_{\rm sp}$. These values of $\xi$ are widely separated; $k_\parallel$ is of the order of $k$ and $|k_{\rm sp}|\gg k$ because $|\varsigma|$ is small. Note that the corresponding values of $Q_+(\xi)$ enter some of the analytical expressions related to the solution $u(x)$, particularly formula~\eqref{eq:u-sp} for the SPP. 

In our computation, we compare approximate formulas~\eqref{eq:Q+_approx-fin} 
and~\eqref{eq:Q+_spec} to a full numerical evaluation of integral~\eqref{eq:Q+_form} for $Q_+(\xi)$. The results are displayed
in Table~\ref{tab:qplus}. Evidently, our approximations are in good agreement
with the numerically computed value of the defining integral. In fact, the relative error is found to be around 2\,\% for
$Q_+(k_\parallel)$ and less than 0.5\,\% for $Q_+(k_{\rm sp})$.

\subsection{Numerically computed solution, $u$}
\label{subsec:num-sol}

\begin{figure}[!t]
  \centering
  \subfloat[Real part]{
    \includegraphics{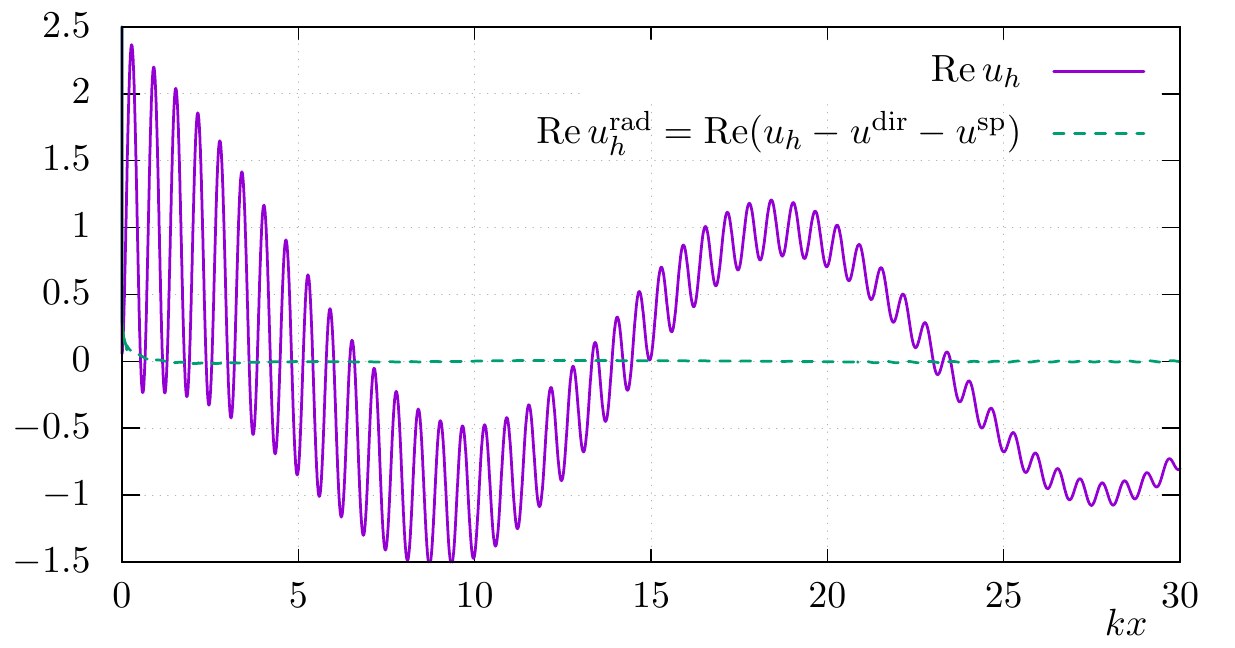}
  }

  \subfloat[Imaginary part]{
    \includegraphics{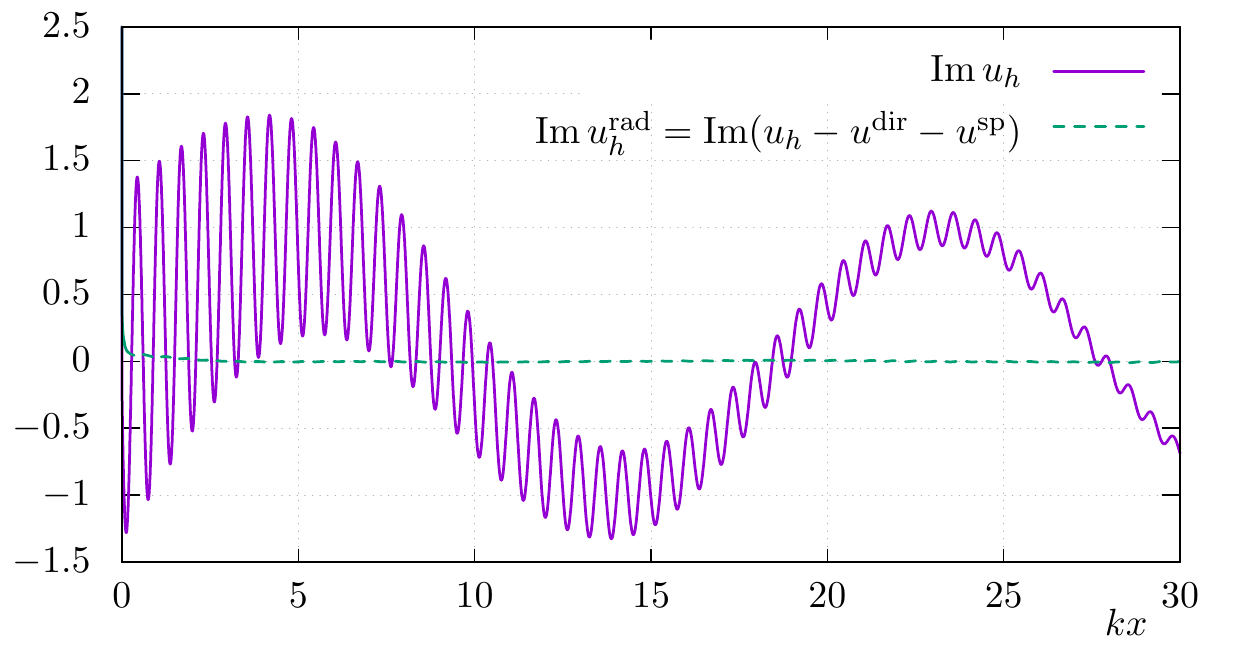}
  }
  \caption{%
    Real part [(a)] and imaginary part [(b)], as a function of $kx$, of:
    the numerically computed solution, $u_h(x)$ (solid line); and the
    corresponding numerical approximation of the radiation field, $u^{\rm
    rad}_h(x) = u_h(x)-u^{\rm dir}(x) - u^{\rm sp}(x)$ (dashes).}
  \label{fig:numerical}
\end{figure}

Next, we perform a finite element computation for the electric field in the presence of the strip in 2D~\cite{MML-jcp}. The computational domain
is chosen to be  large enough to account for distances, $x$, from the edge 
such that $0\le k\,x\le 30$. A numerical approximation, $u_h(x)$, of the tangential ($x$-directed) electric
field, $u(x)$, on the strip is thus extracted from the finite element computation.

The result of this computation for $u(x)$ is shown in Figure~\ref{fig:numerical}.
In the same figure, we also present a plot for the quantity
\begin{align*}
  u^{\rm rad}_h(x):= u_h(x)-u^{\rm dir}(x)-u^{\rm sp}(x)~,
\end{align*}
which pertains to the corresponding approximation of the radiation field, $u^{\rm rad}(x)$. Note that the direct field, $u^{\rm dir}$, is computed by~\eqref{eq:u-dir} and the SPP contribution, $u^{\rm sp}$,
is evaluated by~\eqref{eq:u-sp}. 

A few comments on Figure~\ref{fig:numerical} are in order. 
First, $u_h(x)$ contains a wave with a spatially fast scale, which can be identified with the SPP. Second, it is evident that, for all practical purposes, $u^{\rm rad}_h$ 
has a negligible contribution compared to the SPP and the direct field for $3\le kx\le 30$. Furthermore, the radiation field decays algebraically and approaches zero for increasing and sufficiently large $kx$. This implies that the corresponding diffracted field, 
$u^{\rm df}_h=u_h-u^{\rm dir}$, is dominated by the SPP contribution, $u^{\rm sp}$, unless $x$ lies close enough to the edge ($kx\le 3$); cf.
Sections~\ref{subsec:approx_soln} and~\ref{subsec:rad-nf}.

\subsection{Radiation field near edge: Numerical and analytical solutions}
\label{subsec:rad-nf}

\begin{figure}[!t]
  \centering
  \includegraphics{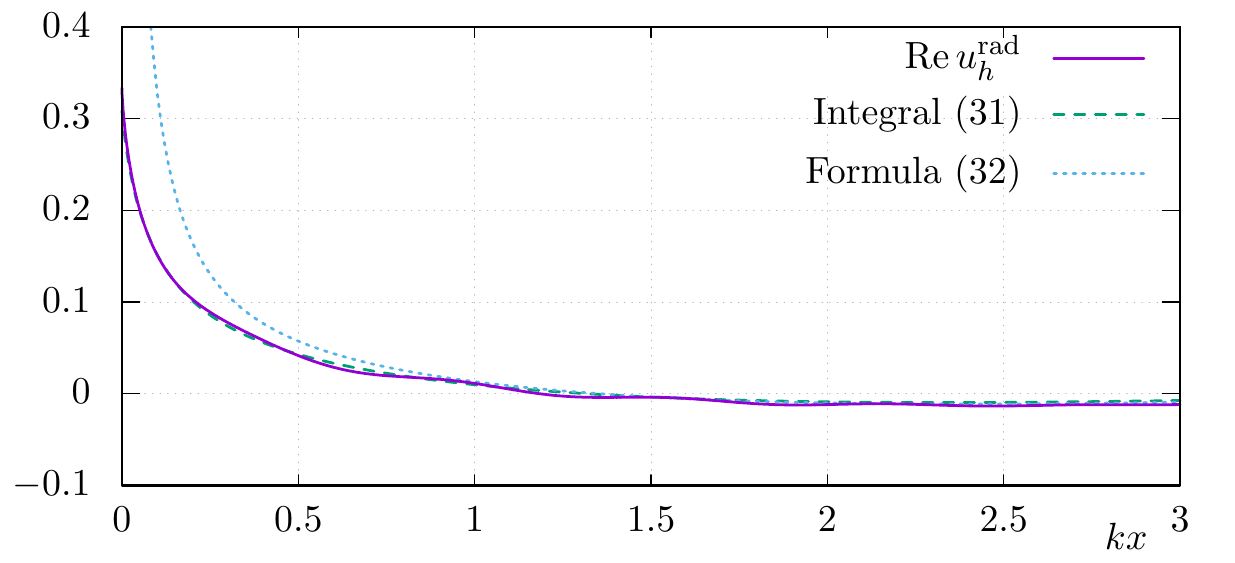}
  \caption{%
    Real part of: numerical approximation, $u^{\rm rad}_h$, of the
    radiation field (solid line); branch-cut integral~\eqref{eq:u-rad-bc}
    (dashes); and approximate formula~\eqref{eq:u-rad-kx} (dotted line).}
  \label{fig:radiation-field}
\end{figure}

In light of the preceding comparison (Section~\ref{subsec:num-sol}), we now restrict attention
to the radiation field, $u^{\rm rad}$, close to the edge of the sheet, for $0\le k\,x\le 3$. In this regime, $u^{\rm rad}(x)$ may be comparable to the diffracted field, $u^{\rm df}(x)$. Accordingly, we carry out another finite element computation in order to obtain a respective numerical resolution, in the proximity of the edge. 

In Figure~\ref{fig:radiation-field}, we indicate the accuracy of our analytical formulas for $u^{\rm rad}$ by showing the real parts of:
(i) the numerically computed radiation field, $u^{\rm rad}_h$;  (ii) the numerically evaluated
integral~\eqref{eq:u-rad-bc} for $u^{\rm rad}$; and (iii) the corresponding approximation
\eqref{eq:u-rad-kx}, valid for $kx=\mathcal{O}(1)$.  A high-order, adaptive
quadrature rule is used to evaluate the requisite integral for $u^{\rm rad}$, while 
$Q_+(\xi)$ is computed with recourse to approximate formula~\eqref{eq:Q+_approx-fin}. 
Evidently, branch-cut integral \eqref{eq:u-rad-bc} is in good
agreement with the finite-element-based result. As we expected, formula
\eqref{eq:u-rad-kx} is reasonably accurate for
$kx\ge1$.

\section{Conclusion}
\label{sec:conclusion}

In this paper, we analyzed a prototypical scattering problem for the excitation by an incident plane wave of a fine-structure SPP on a
semi-infinite metamaterial strip. The governing integral
equation~\eqref{eq:integro-diff} was solved explicitly in terms of a Fourier integral by use of the Wiener-Hopf method. The factorization process for the solution, $u$, was carried out by approximate evaluation of the requisite split function, 
$Q_+(\xi)$, in the Fourier domain.

In our approach, we identified the SPP with a particular
pole contribution to the Fourier representation of the diffracted field; the latter results by substraction from the solution of the incident and directly reflected fields. In addition, by  removal of the SPP from the diffracted field, we derived approximations for
the remaining, radiation field in certain regimes of $kx$.

Our analytical results were found to be in good
agreement with a direct numerical computation based on the finite element
method. In particular, we compared a key integral formula and an ensuing approximate expression for
the radiation field against the finite element, numerical computation. In this vein, we indicated that the radiation field is relatively small for a certain range of values of $kx$ ($kx\ge 2$).

\section*{Acknowledgments}
The first author (DM) is indebted to Professor Tai Tsun Wu for introducing
him to M.~G. Krein's seminal paper~\cite{Krein62}.
The research of the first author (DM) was supported in part by NSF
DMS-1412769.
The research of the second and third authors (MM and ML) was supported in
part by ARO MURI Award W911NF-14-1-0247.

\begin{appendix}

\section{Derivation of integral equation}
\label{app:BVP}
In this appendix, we derive~\eqref{eq:integro-diff} from the appropriate boundary value problem for time-harmonic Maxwell's
equations in the spirit of~\cite{Myers1965}. In 2D, we consider a conducting strip of length $L$, identified with the set $\Omega=\{(x,y): 0\le x\le L,\,y=0\}$; ultimately, let $L\to\infty$. 

For TM polarization, the magnetic field has only a component in the $z$-direction; thus, it can be fully described by a scalar function, $\psi(x,y)$. The vector-valued electric field has components $E_x(x,y)=u(x,y)$ and $E_y(x,y)=v(x,y)$.
The field $(u,v,\psi)$ is decomposed into the incident field, $(u^{\rm in}, v^{\rm in}, \psi^{\rm in})$, and the scattered field, $(u^{\rm s}, v^{\rm s}, \psi^{\rm s})$. The latter solves
\begin{align}
&\frac{\partial \psi^{\rm s}}{\partial y}=-i\frac{k^2}{\omega}u^{\rm s}~,\quad \frac{\partial \psi^{\rm s}}{\partial x}=i\frac{k^2}{\omega}v^{\rm s}~,\label{eq:Amp-Maxw}\\
&\frac{\partial v^{\rm s}}{\partial x}-\frac{\partial u^{\rm s}}{\partial y}=i\omega \psi^{\rm s}~;\quad (x,y)\in\mathbb{R}^2\setminus \Omega~.\label{eq:Faraday}
\end{align}
Thus, $\psi^{\rm s}$ satisfies the Helmholtz equation outside $\Omega$, viz.,
\begin{equation}\label{eq:psi-Helmh}
\Delta_{x,y}\psi^{\rm s}+k^2\psi^{\rm s}=0~,\quad (x,y)\in\mathbb{R}^2\setminus\Omega~.
\end{equation}
In addition, $\psi^{\rm s}$ obeys the following boundary conditions~\cite{Bludov13}:
\begin{equation}\label{eq:BCs-sheet}
\biggl[\frac{\partial \psi^{\rm s}}{\partial y}\biggr]_\Omega=\frac{k^2}{i\omega}[u^{\rm s}]_\Omega=0~,\quad
[\psi^{\rm s}]_\Omega=[\psi]_\Omega=\mu\sigma\, u=\frac{i\omega\mu\sigma}{k^2}\frac{\partial\psi}{\partial y}\quad \mbox{on}\ \Omega~;
\end{equation}
and the Sommerfeld radiation condition as $\sqrt{x^2+y^2}\to\infty$. Here, $\psi=\psi^{\rm s}+\psi^{\rm in}$; $[Q]_\Omega=Q|_{y=0^+}-Q|_{y=0^-}$ denotes the jump of $Q(x,y)$ across $\Omega$ ($0< x<L$); $\sigma$ is the (constant) conductivity of the sheet; and $\mu$ is the magnetic permeability of the ambient medium. We also require that $\partial\psi/\partial y$ be bounded at $x=0,\,L$.
The external source that produces $(u^{\rm in}, v^{\rm in}, \psi^{\rm in})$ lies outside $\Omega$; thus, the incident field is smooth in a neighborhood of $\Omega$.

At this stage, we choose to proceed by converting the boundary value problem for $\psi^{\rm s}$ into an integral equation for $u$. To this end, consider the Green function, $G(x',y';x,y)$, for the Helmholtz equation on $\mathbb{R}^2$ defined via
\begin{equation}\label{eq:Green-pde}
(\Delta_{x,y}+k^2)G=-\delta(x-x')\delta(y-y')~;\quad (x,y)\in \mathbb{R}^2~,\ (x',y')\in \mathbb{R}^2\setminus\Omega~,
\end{equation}
and imposition of the Sommerfeld radiation condition at infinity. For fixed $(x',y')$ not in $\Omega$, this $G(x',y';\cdot)$ is infinitely differentiable in a neighborhood of $\Omega$. Specifically, note that~\cite{CourantHilbert}
\begin{equation}\label{eq:Green-form}
G(x',y';x,y)=\frac{i}{4}H_0^{(1)}(k\sqrt{(x-x')^2+(y-y')^2})~,
\end{equation}
which has the Fourier transform
\begin{equation}\label{eq:Green-FT}
\widehat{G}(\xi,y;0,0)=\int_{-\infty}^{\infty}G(x,y;0,0)e^{-i\xi x}\,\db x=\frac{i}{2\sqrt{k^2-\xi^2}}e^{i\sqrt{k^2-\xi^2}|y|}~,
\end{equation}
with $\I \sqrt{k^2-\xi^2}>0$. The last inequality is dictated by the Sommerfeld radiation condition, and defines the top Riemann sheet.

By a standard procedure of scattering theory~\cite{Myers1965,Colton83}, a combination of \eqref{eq:psi-Helmh} and~\eqref{eq:Green-pde} is suitably integrated on $\mathbb{R}^2\setminus\Omega$. The application of the divergence theorem then yields the integral relation (with primed and unprimed coordinates interchanged)
\begin{equation*}
\psi^{\rm s}(x,y)=\int_0^L \frac{\partial G(x,y;x',y')}{\partial y'}\biggr|_{y'=0}\,[\psi^{\rm s}(x',y'=0^+)-\psi^{\rm s}(x',y'=0^-]\,\db x'~,
\end{equation*}
where $(x,y)\in \mathbb{R}^2\setminus\Omega$ and use was made of the continuity of $\partial\psi^{\rm s}/\partial y$ across $\Omega$. By $\partial G/\partial y'=-\partial G/\partial y$ and the second condition of~\eqref{eq:BCs-sheet}, we have
\begin{equation*}
\psi^{\rm s}(x,y)=-\mu\sigma\frac{\partial}{\partial y}\int_0^L G(x,y;x',0)\,u(x',0)\,\db x'~,\quad (x,y)\notin\Omega~.
\end{equation*}
Now differentiate both sides of the last equation with respect to $y$ to obtain
\begin{equation*}
u^{\rm s}(x,y)=\frac{i\omega}{k^2}\frac{\partial}{\partial y}\psi^{\rm s}(x,y)=-\frac{i\omega\mu\sigma}{k^2}\frac{\partial^2}{\partial y^2}\int_0^L G(x,y;x',0)\,u(x',0)\,\db x'~.
\end{equation*}
By using the equation
\begin{equation*}
(\Delta_{x,y}+k^2)\int_0^L G(x,y;x',0)\,u(x',0)\,\db x'=0~,\quad (x,y)\in \mathbb{R}^2\setminus\Omega~,
\end{equation*}
and then allowing $(x,y)$ to approach the strip, $\Omega$, we find that $u$ satisfies
\begin{equation*}
u(x,0)-u^{\rm in}(x,0)=\frac{i\omega\mu\sigma}{k^2}\biggl(\frac{\partial^2}{\partial x^2}+k^2\biggr)\int_0^L G(x-x',0;0,0)\,u(x',0)\,\db x'~,
\end{equation*}
where $0< x< L$.
Equation~\eqref{eq:integro-diff} is thus recovered as $L\to\infty$, with $\mathcal K(x):=G(x,0;0,0)$. By slightly abusing notation, we denote $u(x,0)$ by $u(x)$.

\setcounter{equation}{0}

\section{Wiener-Hopf method: A review}
\label{app:WH}
In this appendix, we formally outline the core ideas of the Wiener-Hopf method in some correspondence to~\cite{Krein62,Masujima-book}. Although the starting point is~(\ref{eq:WH-IE}), the factorization method can treat integral equation~\eqref{eq:integro-diff} on the basis of generic functional equation~(\ref{eq:WE-u-FT}).

Consider the second-kind Wiener-Hopf integral equation
\begin{equation}\label{eq:WH-IE}
  u(x)-\lambda \int_0^\infty
  K(x-y)\,u(y)\,\db y=f(x)~,\quad x>0~,
\end{equation}
where $f(x)$ is given and $u(x)$ must
be determined; and $\lambda$ is a given, complex parameter.\footnote{In integral equation~\eqref{eq:integro-diff}, this $\lambda$ is replaced by the 1D Helmholtz operator. However, the main steps of the Wiener-Hopf method described here remain essentially intact.} Suppose that $u(x)$ and $f(x)$ are integrable on $[0, \infty)$. If
$K(x)$ is integrable in $(-\infty, \infty)$, it has a continuous, single-valued Fourier transform,
$\widehat{K}(\xi)$, for all real $\xi$. More generally, let us write
\begin{displaymath}
  K(x)=\frac{1}{2\pi}\int_{-\infty}^\infty
  \widehat{K}(\xi)\,e^{i\xi x}\,\db \xi~,
\end{displaymath}
assuming that $\widehat{K}(\xi)$ is holomorphic in
a region of the $\xi$-plane that contains the real axis.

Next, we extend the domain of~(\ref{eq:WH-IE}) to the whole real axis;
thus, define
\begin{equation}
  u(x)\equiv 0\quad\mbox{and}\quad f(x)\equiv 0\quad \mbox{if}\quad x<0~.
\end{equation}
Accordingly, (\ref{eq:WH-IE}) reads
\begin{equation}\label{eq:WH-IE-ex}
  u(x)-\lambda\int_{-\infty}^\infty K(x-y)\,u(y)\,\db y=f(x)+g(x)~,\quad
  x\in\mathbb{R}~,
\end{equation}
where
\begin{equation}\label{eq:g-def}
  g(x):=\left\{ \begin{array}{lr} -\lambda \int_0^\infty
  K(x-y)\,u(y)\,\db y~, & x<0~,\cr 0~, & x>0~. \end{array} \right.
\end{equation}
Evidently, this $g$ is unknown.

We define the Fourier transform, $\widehat{u}$, of $u$ by
\begin{equation}\label{eq:u-FT}
  \widehat{u}(\xi)=\int_{-\infty}^\infty u(x)\, e^{-i\xi
  x}\,\db x=\int_0^\infty u(x)\,e^{-i\xi x}\,\db x~,\quad \xi\in\mathbb{R}~.
\end{equation}
By our assumption for $u(x)$, especially its vanishing for $x<0$, this $\widehat{u}(\xi)$ can be viewed as the analytic continuation to the real axis of a holomorphic function in the lower half-plane,
$\mathbb{C}_-=\{\xi\in\mathbb{C}\,:\,\text{Im}\xi< 0\}$. We write
\begin{equation}\label{eq:InvFT-u}
  u(x)=\int_{\Gamma}\widehat{u}(\xi)\,e^{i\xi x}\ \frac{\db\xi}{2\pi}~;\
  \Gamma=\{\xi\in \mathbb{C}\,:\, \text{Im}\xi={\rm const.}< 0\}~.
\end{equation}
In a similar vein, the Fourier
transform, $\widehat{f}(\xi)$, of $f$ is holomorphic in
$\mathbb{C}_-$. In regard to $g$, define\looseness=-1
\begin{equation}\label{eq:u-FT}
  \widehat{g}(\xi)=\int_{-\infty}^\infty g(x)\, e^{-i\xi
  x}\,\db x=\int_{-\infty}^0 g(x)\,e^{-i\xi x}\,\db x~.
\end{equation}
If $g(x)$ is integrable in $(-\infty, 0]$ then \eqref{eq:u-FT} is defined in the upper half-plane,
$\mathbb{C}_+=\{\xi\in\mathbb{C}\,:\,\text{Im}\xi> 0\}$, where
$\widehat{g}(\xi)$ is holomorphic; cf.~(\ref{eq:g-def}). We analytically continue $\widehat{g}(\xi)$ from $\mathbb{C}_+$ to the real axis.

The Fourier transformation of~(\ref{eq:WH-IE-ex}) yields the functional equation
\begin{equation}\label{eq:WE-u-FT}
  \bigl [1-\lambda
  \widehat{K}(\xi)\bigr]\,\widehat{u}(\xi)=\widehat{f}(\xi)+\widehat{g}(\xi)~,
\end{equation}
where
$\xi\in\mathbb{R}\setminus\{\xi\in\mathbb{R}\,:\,\widehat{f}(\xi)=\infty\
\mbox{or}\ \widehat{g}(\xi)=\infty\}$. We assume that
$\lambda$ is such that
\begin{displaymath}
  1-\lambda \widehat{K}(\xi)\neq 0\quad \mbox{for\ all\ real}\ \xi~;
\end{displaymath}
also, we do not allow real limit points of complex zeros of $1-\lambda
\widehat{K}(\xi)$. Although we derived~(\ref{eq:WE-u-FT})
from~(\ref{eq:WH-IE}), it is important to consider~(\ref{eq:WE-u-FT})
as a functional equation that may be a starting point in its own right. Notably, (\ref{eq:WE-u-FT}) can result directly from a class of linear
boundary value problems~\cite{Noble-book}.

Next, we focus on determining $\widehat{u}(\xi)$ from~(\ref{eq:WE-u-FT}).
The primary task is to explicitly separate~(\ref{eq:WE-u-FT}) into
two parts, one of which is holomorphic in $\mathbb{C}_+$ and another that is
holomorphic in $\mathbb{C}_-$. For this purpose, define~\cite{Krein62}
\begin{equation}\label{eq:Q-def}
  \widehat{Q}(\xi)=\ln \bigl[1-\lambda \widehat{K}(\xi)\bigr]~.
\end{equation}
This $\widehat{Q}(\xi)$ is single valued and holomorphic in a region of the complex plane
that includes the real axis.
We need to find $\widehat{Q}_\pm(\xi)$ such that
\begin{equation}\label{eq:Q_+-sum}
  \widehat{Q}(\xi)=\widehat{Q}_+(\xi)+\widehat{Q}_-(\xi)~,\quad \xi\in\mathbb{R}~,
\end{equation}
where $\widehat{Q}_s(\xi)$ is holomorphic in $\mathbb{C}_s$ ($s=\pm$). We refer to each $Q_\pm(\xi)$ as a `$\pm$' split function.

We slightly digress to discuss particulars of~\eqref{eq:Q-def} and introduce the notion of the {\em index} from Krein's theory~\cite{Krein62}. Bearing in mind the role of $K$ in $\widehat{Q}$ above, we expect that
a continuous and single-valued branch of
$\widehat{Q}(\xi)$ can be chosen such that
$\arg[1-\lambda\widehat{K}(\xi)]-\arg[1-\lambda\widehat{K}(-\xi)]\to 0$ as
$\xi\to +\infty$. This amounts to a zero {\em index},
${\rm ind}(1-\lambda \widehat{K})=0$, by the following
definition~\cite{Krein62}.

\begin{definition}[Index of functional equation]\label{def:index}
The {\rm index} of~\eqref{eq:WE-u-FT} is
\begin{equation}\label{eq:index-eq}
  {\rm ind}(1-\lambda \widehat{K}):=\frac{1}{2\pi}\big[\arg
  \big(1-\lambda\widehat{K}(\xi)\big)\big]\bigl|_{\xi=-\infty}^\infty~.
\end{equation}
\end{definition}

We mention in passing a rigorous result from~\cite{Krein62} in relation to this index. Let $\mathcal E$ denote a functional space from a family of spaces that includes, e.g., any space $L^p(0,\infty)$ with $p\ge 1$ and the space of all bounded continuous functions on $(0,\infty)$~\cite{Krein62}. Note the following excerpt of a theorem in~\cite{Krein62}.

\begin{theorem}\label{thm:Krein-theorem}
Let $K(x)$ be integrable in $(-\infty, \infty)$. Then,~\eqref{eq:WH-IE} has exactly one solution $u\in \mathcal E$ for an arbitrary $f\in \mathcal E$ if and only if the conditions
\begin{equation}\label{eq:conds-K}
1-\lambda\widehat{K}(\xi)\neq 0\quad (\xi\in\mathbb{R})\quad \mbox{\rm and}\quad {\rm ind}(1-\lambda\widehat{K})=0
\end{equation}
are satisfied.
\end{theorem}

\begin{figure}
  \begin{center}
  \includegraphics*[scale=0.5, trim=0in 2.0in 0in 1.4in]{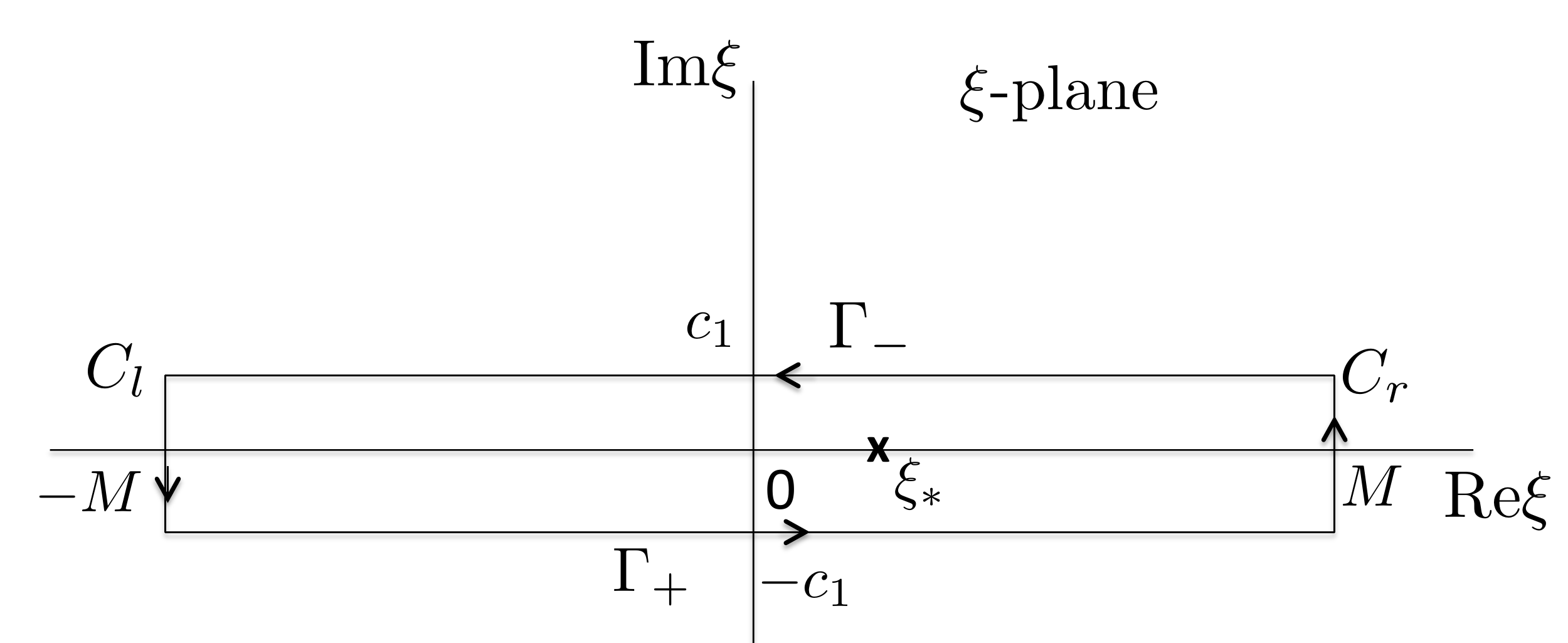}
  \end{center}
  \caption{Contour of integration, $\Gamma=\Gamma_+ \cup \Gamma_- \cup C_l
    \cup C_r$, for $Q_s(\xi)$ ($s=\pm$). The arrows indicate sense of circulation. The oriented path
    $\Gamma_s=\{\xi\in\mathbb{C}\,:\,\text{Im}\xi=-sc_1~,\, -M
    < \text{Re}\xi < M\}$, for small $c_1$, lies in $\mathbb{C}_s$ ($s=\pm $). The path
    $C_{p}=\{\xi\in\mathbb{C}\,:\,\text{Re}\xi=\mp M~,\,
    |\text{Im}\xi|< c_1\}$ ($p=l,\,r$) is a small segment connecting
    $\Gamma_+$ and $\Gamma_-$ on the left ($p=l$) or right ($p=r$) of the
    imaginary axis; eventually,
    let $M\to +\infty$.}
  \label{fig:figA1}
\end{figure}

Henceforth, we assume that~\eqref{eq:conds-K} hold. We return to~\eqref{eq:Q_+-sum}. To determine $Q_s$ ($s=\pm$), consider the contour $\Gamma=\Gamma_+ \cup \Gamma_- \cup C_p \cup
C_l$, as shown in Figure~\ref{fig:figA1}. Let
$\xi_*\in\mathbb{R}$ be an arbitrary point enclosed by $\Gamma$. By the
Cauchy integral formula and the property ${\rm ind}(1-\lambda\widehat{K})=0$, we write
\begin{eqnarray}\label{eq:Cauchy-IF}
  \widehat{Q}(\xi_*)&=&\frac{1}{2\pi i}\lim_{M\to+\infty}\left(\int_{\Gamma_+(M)}
  +\int_{\Gamma_-(M)}+\int_{C_l(M)}+\int_{C_r(M)}\right)\frac{\widehat{Q}(\xi)}{\xi-\xi_*}\ \db\xi
  \nonumber\\
  \mbox{} &=& \frac{1}{2\pi i}\int_{\Gamma_+(\infty)}
  \frac{\widehat{Q}(\xi)}{\xi-\xi_*}\ \db\xi+\left(-\frac{1}{2\pi
  i}\int_{-\Gamma_-(\infty)}
  \frac{\widehat{Q}(\xi)}{\xi-\xi_*}\ \db\xi\right)
  \nonumber\\
  \mbox{}&&\quad
  +\frac{1}{2\pi}\lim_{M\to\infty}\int_{-c_1}^{c_1}
  \left(\frac{\widehat{Q}(\xi)}{\xi-\xi_*}-\frac{\widehat{Q}(-\xi)}{-\xi-\xi_*}
  \right)\Biggl|_{\xi=M+i\varrho}\ \db\varrho~,
\end{eqnarray}
where the integrals in the second line are presumed convergent; cf.~\eqref{eq:Qs-int}. In regard to the third line
of~(\ref{eq:Cauchy-IF}), we assert that the integral
approaches zero as $M\to \infty$ if the growth of $\widehat{K}(\xi)$ is not faster than polynomial ($\xi=M+i\varrho$, $|\varrho|\le c_1$). Now consider each of the integrals in the
second line of~(\ref{eq:Cauchy-IF}) as a function of $\xi_*$; and let
$\xi_*$ be moved slightly off the real axis, to $\mathbb{C}_+$ or $\mathbb{C}_-$.
Each of these integrals is then carried out on the real axis, viz.,
\begin{equation*}
  \int_{\Gamma_s(\infty)}\db\xi\ (\cdot)=\lim_{M\to +\infty}\int_{-M}^M
  \db\xi\ (\cdot)~,\quad \mbox{if}\quad s\,\text{Im}\xi_*> 0~.
\end{equation*}
Hence, we deduce that
\begin{align}\label{eq:Qs-int}
  \widehat{Q}_s(\xi_*)&=s\,\frac{1}{2\pi i}\lim_{M\to\infty}\int_{-M}^M
  \db\xi\ \frac{\widehat{Q}(\xi)}{\xi-\xi_*}\qquad (s=\pm)~,\notag\\
  &=s\frac{1}{2\pi i}\lim_{M\to\infty}\int_0^M \biggl[\frac{\widehat{Q}(\xi)}{\xi-\xi_*}-\frac{\widehat{Q}(-\xi)}{\xi+\xi_*}\biggr]\ \db\xi~,
\end{align}
where $s\,\text{Im}\xi_*>0$. Each of these two integrals ($s=\pm$) converges if $\widehat{K}(\xi)$ has at most polynomial growth. In fact, the condition ${\rm ind}(1-\lambda\widehat{K})=0$ guarantees that $\widehat{Q}(\xi)-\widehat{Q}(-\xi)$ approaches 0 as $\xi\to +\infty$.

Accordingly,~(\ref{eq:WE-u-FT}) is recast into
\begin{equation}\label{eq:WE-u-FT-mod}
  e^{\widehat{Q}_-(\xi)}\,\widehat{u}(\xi)=e^{-\widehat{Q}_+(\xi)}\widehat{f}
  (\xi)+e^{-\widehat{Q}_+(\xi)}\widehat{g}(\xi)~.
\end{equation}
The left-hand side is holomorphic in $\mathbb{C}_-$, while the second term on the
right-hand side is holomorphic in $\mathbb{C}_+$. By further decomposing
$e^{-\widehat{Q}_+(\xi)}\widehat{f}(\xi)$ as
\begin{equation*}
  e^{-\widehat{Q}_+(\xi)}\widehat{f}(\xi)=\biggl[e^{-\widehat{Q}_+(\xi)}
  \widehat{f}(\xi)\biggr]_+ + \biggl[e^{-\widehat{Q}_+(\xi)}\widehat{f}(\xi)
  \biggr]_-~,
\end{equation*}
we derive the expression
\begin{equation}\label{eq:WE-u-FT-pm}
  e^{\widehat{Q}_-(\xi)}\,\widehat{u}(\xi)-\biggl[e^{-\widehat{Q}_+(\xi)}
  \widehat{f}(\xi)\biggr]_-=\biggl[e^{-\widehat{Q}_+(\xi)}\widehat{f}(\xi)
  \biggr]_+ +e^{-\widehat{Q}_+(\xi)}\widehat{g}(\xi)~,
\end{equation}
where
\begin{equation}\label{eq:expQ_s}
  \biggl[e^{-\widehat{Q}_+(\xi)}\widehat{f}(\xi)\biggr]_s=s\frac{1}{2\pi
  i}\int_{-\infty}^{\infty}\db\zeta\ \frac{e^{-\widehat{Q}_+(\zeta)}
  \widehat{f}(\zeta)}{\zeta-\xi}\qquad (s=\pm)
\end{equation}
with $s\,\text{Im} \xi>0$, under the assumption that these integrals converge.\footnote{If
$\widehat{f}(\zeta)$ has singularities on the real axis, the path of
integration in~(\ref{eq:expQ_s}) is indented below ($s=-$) or above ($s=+$) these
singularities, with the point $\xi$ keeping its position {\em relative} to
the path.}

In~(\ref{eq:WE-u-FT-pm}), the left-hand side defines a function holomorphic
in $\mathbb{C}_-$ while the right-hand side is holomorphic in $\mathbb{C}_+$; these two
functions are equal on the real axis. Hence, these functions together define an {\em entire} function,
$E(\xi)$. In particular, we have
\begin{equation}\label{eq:E-lo}
  e^{\widehat{Q}_-(\xi)}\,\widehat{u}(\xi)-\biggl[e^{-\widehat{Q}_+(\xi)}
  \widehat{f}(\xi)\biggr]_-=E(\xi)~,\quad \xi\in \mathbb{C}_-~,
\end{equation}
and
\begin{equation}\label{eq:E-ri}
  \biggl[e^{-\widehat{Q}_+(\xi)}\widehat{f}(\xi)\biggr]_+
  +e^{-\widehat{Q}_+(\xi)}\widehat{g}(\xi)=E(\xi)~,\quad \xi\in\mathbb{C}_+~.
\end{equation}
Thus, we conclude that
\begin{equation}\label{eq:u-FT-fin}
  \widehat{u}(\xi)=e^{-\widehat{Q}_-(\xi)}\left\{
  \biggl[e^{-\widehat{Q}_+(\xi)}\widehat{f}(\xi)\biggr]_-
  +E(\xi)\right\}~,\quad \xi\in \mathbb{C}_-~.
\end{equation}
The last formula can be Fourier-inverted to yield $u(x)$ according
to~(\ref{eq:InvFT-u}).

The function $E(\xi)$ can in principle be determined by its behavior in
$\mathbb{C}_\pm$ as $\xi\to\infty$. This behavior may be extracted
from~(\ref{eq:E-lo}) and~(\ref{eq:E-ri}). If the left-hand sides of these
equations approach $0$ as $\xi\to\infty$ in $\mathbb{C}_-$ and
$\mathbb{C}_+$, then, by Liouville's theorem, $E(\xi)\equiv 0$. This result
is consistent with the existence of a unique solution $u$ in the case of
Theorem~\ref{thm:Krein-theorem}.

\end{appendix}

\bibliographystyle{sapm}

\end{document}